\documentclass[prl,superscriptaddress,twocolumn,floatfix]{revtex4}
\usepackage{graphicx}

\def\lsco{La$_{2-x}$Sr$_x$CuO$_4$}
\def\lbco{La$_{2-x}$Ba$_x$CuO$_4$}
\def\lbcoate{La$_{1.875}$Ba$_{0.125}$CuO$_4$}
\def\lnsco{La$_{1.6-x}$Nd$_{0.4}$Sr$_x$CuO$_4$}

\def\tso{$T_{\rm so}$}
\def\tco{$T_{\rm co}$}
\def\tctd{$T_c^{\rm 2D}$}
\def\tbkt{$T_{\rm BKT}$}

\begin{document}

\title{Two-Dimensional Superconducting Fluctuations in Stripe-Ordered
\lbcoate}
\author{Q. Li}
\author{M. H\"ucker}
\author{G. D. Gu}
\author{A. M. Tsvelik}
\author{J. M. Tranquada}
\affiliation{Brookhaven National Laboratory, Upton, NY 11973-5000} 
\date{\today}
\begin{abstract}
Recent spectroscopic observations of a $d$-wave-like gap in stripe-ordered
\lbco\ with $x=\frac18$ have led us to critically analyze the anisotropic
transport and magnetization properties of this material.  The data
suggest that concomitant with the spin ordering is an electronic
decoupling of the CuO$_2$ planes.  We observe a transition (or crossover)
to a state of two-dimensional (2D) fluctuating superconductivity, which
eventually reaches a 2D superconducting state below a
Berezinskii-Kosterlitz-Thouless transition.  Thus, it appears that the
stripe order in \lbco\ frustrates three-dimensional superconducting phase
order, but is fully compatible with 2D superconductivity and an enhanced
$T_c$.
\end{abstract}
\maketitle

Charge and spin stripe order have been observed experimentally in a few
special cuprate compounds, specifically \lbco\ \cite{fuji04} and \lnsco\
\cite{ichi00}.  Some theoretical studies have proposed that stripe
correlations should be good for pairing and high superconducting
transition temperatures, $T_c$ \cite{emer97}; however, such notions have
been highly controversial, given that the highest stripe ordering
temperatures occur at $x=\frac18$, where $T_c$ is strongly depressed. 
A recent study of \lbcoate\ with angle-resolved photoemission and
scanning tunneling spectroscopies (STS) \cite{vall06} has found evidence
for a $d$-wave-like gap at low temperature, well within the stripe-ordered
phase but above the bulk $T_c$. An earlier infrared reflectivity study
\cite{home06} demonstrated that an anisotropic gap, together with a
narrowed Drude component, becomes apparent as soon as one cools below the
charge-ordering temperature, $T_{\rm co}=54$~K.  Is the observed gap due
to exotic electron-hole pairing that reduces the density of states
available for the formation of Cooper pairs?  Alternatively, could the
gap be associated with particle-particle pairing, but with stripe order
interfering with superconducting phase order?  In an attempt to
resolve this issue, we have carefully studied the anisotropic transport
and magnetization properties of \lbcoate.

\begin{figure}
\centerline{\includegraphics[width=3.2in]{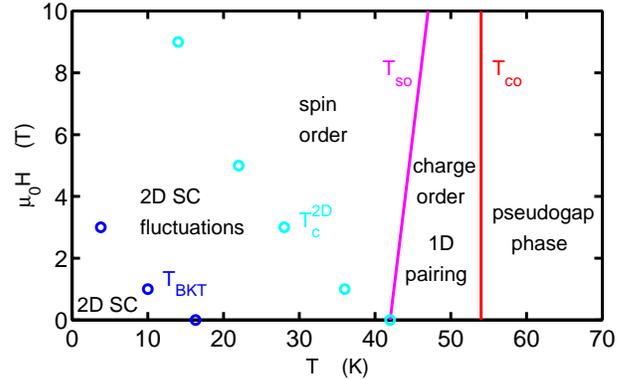}}
\caption{(color online)  Experimental phase diagram for \lbcoate.  The
transition lines for charge order and spin order are from \cite{huck05b}. 
The boundaries labeled \tctd\ and \tbkt\ are described in the text.}
\label{fg:phase_diag} 
\end{figure}

In this Letter, we present compelling evidence that the dominant impact of
the stripe ordering is to electronically decouple the CuO$_2$ planes. 
The charge-ordering transition, at \tco, is correlated with a rapid
increase in the anisotropy between the resistivity along the $c$-axis,
$\rho_c$, and that parallel to the CuO$_2$ planes, $\rho_{ab}$.  At the
spin-ordering temperature, \tso, there is a sharp drop in $\rho_{ab}$ by
an order of magnitude; we label the latter magnetic-field-dependent
transition as \tctd (see Fig.~\ref{fg:phase_diag}).  Below \tctd,
$\rho_{ab}(T)$ follows the temperature dependence predicted \cite{halp79}
for a 2D superconductor above its Berezinskii-Kosterlitz-Thouless
transition temperature, \tbkt\
\cite{bere71,kost73}.  This state also exhibits weak, anisotropic
diamagnetism and a thermopower very close to zero.  Below the nominal
\tbkt ($\sim16$~K), we observe nonlinear voltage-current ($V$--$I$)
behavior consistent with expectations for a 2D superconductor
\cite{minn87}.  We conclude that charge inhomogeneity and 1D correlations
are good for pairing in the CuO$_2$ planes, as has been argued
theoretically \cite{emer97,tsve07}; however, the interlayer Josephson
coupling is effectively zero in the stripe-ordered state of \lbcoate.

The crystals studied here were grown in an infrared image furnace by the
floating-zone technique.  They are pieces from the same crystals used
previously to characterize the optical conductivity \cite{home06},
photoemission and STS \cite{vall06}, magnetization \cite{huck05b}, and
magnetic excitations
\cite{tran04}.  In particular, the charge-stripe order has been
characterized by soft x-ray resonant diffraction \cite{abba05} and by
diffraction with 100-keV x-rays \cite{huckun}.  The latter results show
that $T_{\rm co}$ occurs at precisely the same temperature as the
structural phase transition, from orthorhombic ($Bmab$) to 
tetragonal ($P4_2/ncm$) symmetry.  (Note that the structural transition
is first order, with a two-phase coexistence region extending over a
couple of degrees.) The spin ordering of the stripes, as determined by
neutron diffraction \cite{fuji04}, muon spin rotation spectroscopy
\cite{savi05}, and high-field susceptibility
\cite{huck05b}, occurs at $\sim40$~K \cite{LBCO_note}.

\begin{figure}
\centerline{\includegraphics[width=2.8in]{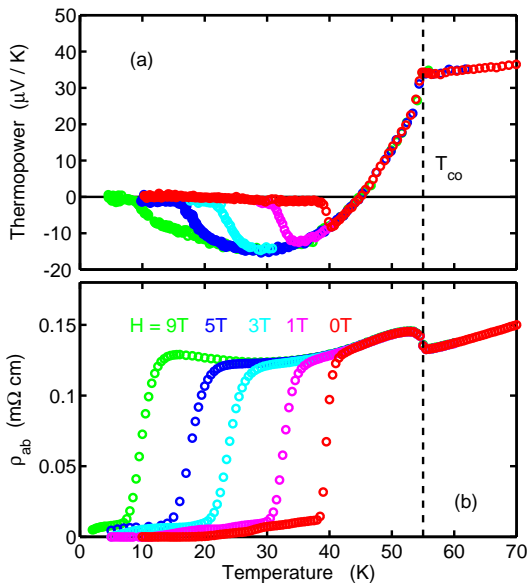}}
\caption{(color online) (a) Thermoelectric power vs.\ temperature for
several different magnetic fields (as labeled in (b)), applied along the
$c$-axis.  (b) In-plane resistivity vs.\ temperature for the same magnetic
fields as in (a).  The vertical dashed line indicates $T_{\rm co}$.}
\label{fg:S_rho} 
\end{figure}

Transport measurements were carried out by the four-probe method on two
single crystals cut side-by-side from the same slab.  The parent slab
exhibited a bulk diamagnetic transition at 4~K, with 100\%\ magnetic
shielding at lower temperatures.  To measure 
$\rho_{ab}$, current contacts were made at the ends of the long
crystal ($7.5\times2$ mm$^2\times 0.3$ mm along the $c$-axis) to ensure
uniform current flow; voltage pads were also in direct contact with the
$ab$-plane edges.  Voltage-current characteristics were measured over 5
orders of magnitude with pulsed current ($\le1$~ms) to avoid sample
heating. The thermoelectric power was measured by the four-probe dc
steady state method with a temperature gradient along the $ab$-plane at
1\%\ of $T$ across the crystal. For $\rho_c$, current contacts covered the
major part (85\%) of the broad surfaces of the crystal ($7.5\times3.4$
mm$^2\times1.15$ mm along $c$) to ensure uniform current flow, with
voltage contacts on the same surfaces, occupying 5\%\ of the area. By
annealing the contact pads (Ag paint) at 200--450$^\circ$C for 0.5~h
under flowing O$_2$, low contact resistance ($<0.2$ $\Omega$) was always
obtained. Annealing the crystals under flowing O$_2$ at 450$^\circ$C for
100~h did not alter the transport results.  All resistivity data reported
here were taken with a dc current of 5 mA.  The
magnetic susceptibility was measured on a third crystal, having a mass of
0.6~g using a SQUID (superconducting quantum interference device)
magnetometer.

\begin{figure}[t]
\centerline{\includegraphics[width=3.2in]{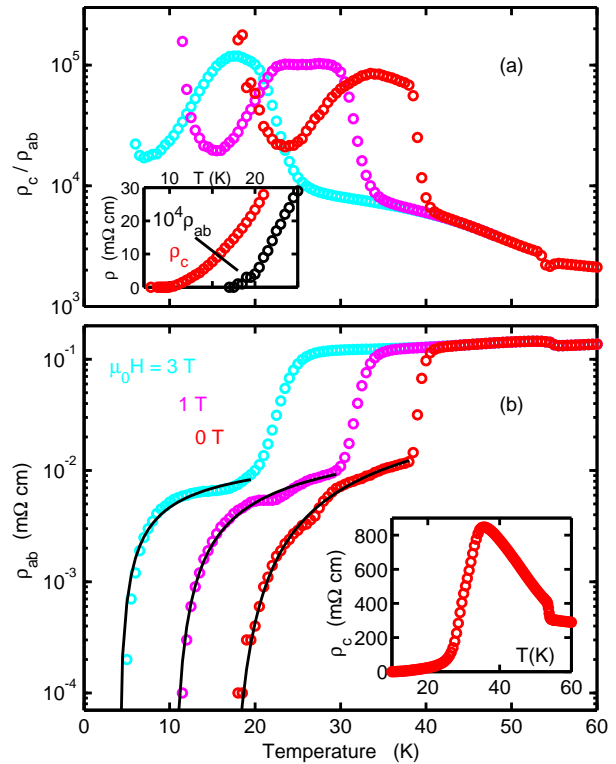}}
\caption{(color online) (a) Ratio of $\rho_c$ to $\rho_{ab}$ vs.\
temperature in fields of 0, 1~T, and 3~T, as labeled in (b).  Inset shows
resistivity vs.\ temperature; note that $\rho_{ab}$
reaches zero (within error) at 18~K, while $\rho_c$ does not reach zero
until 10~K.  (b) In-plane resistivity vs.\ temperature on a semilog
scale, for three different
$c$-axis magnetic fields, as labeled.  The lines through the data points
correspond to fits to Eq.~(1). Inset shows $\rho_c$ at zero
field on a linear scale.}
\label{fg:log_rho} 
\end{figure}

Let us first consider the changes near \tco. Figure 2 shows the
thermopower and $\rho_{ab}$ as a function of
temperature. The thermopower shows a drastic drop below the transition,
going slightly negative below 45~K.  This behavior is consistent with
previous studies of the thermopower and Hall effect in
La$_{2-x-y}$Nd$_y$Sr$_x$CuO$_4$ and \lbco\ \cite{huck98,noda99,adac01}. 
In contrast, $\rho_{ab}$ shows a modest jump and then continues downward
with a slope similar to that above the transition; the sheet resistance
at 45~K is $\sim2$~k$\Omega$, well within the metallic regime.  Consider
also the results for
$\rho_c/\rho_{ab}$, shown in Fig.~\ref{fg:log_rho}(a).  
This ratio grows on cooling, especially below \tco; such behavior is
inconsistent with expectations for a Fermi-liquid.

The drop in thermopower suggests that the densities of filled and empty
states close to the Fermi level become more symmetric when charge-stripe
order is present.  At the same time, the small change in $\rho_{ab}$
indicates that the dc conductivity in the planes remains essentially
2D.  We also know that the gap feature in the optical conductivity
shows up below \tco\ \cite{home06}.  Since the gap does not seem to
impact the 2D conductivity, it appears that it must be associated with 1D
correlations within the stripes \cite{emer97,tsve07,pera96}.  A possible
model for this state is the ``sliding'' Luttinger-liquid phase
\cite{emer00}, especially in the form worked out for neighboring layers
of orthogonal stripes
\cite{mukh01}, since we know that the orientation of the charge stripes
rotates by
$\pi/2$ from one layer to the next, following the glide symmetry of
the crystal structure \cite{vonz98}.  The latter model predicts both 2D
metallic resistivity in the planes and
$\rho_{c}/\rho_{ab}\sim T^{-\alpha}$ with $\alpha>1$, qualitatively
consistent with our observations.

\begin{figure}[t]
\centerline{\includegraphics[width=2.8in]{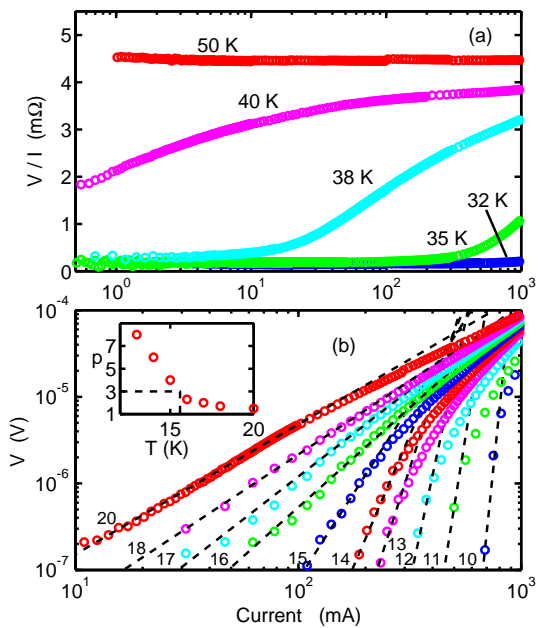}}
\caption{(color online) (a) In-plane $V/I$ vs.\ $I$ at
five temperatures, as labeled. Note that $I$ is on a log scale. (b)
Log-log plot of in-plane $V$ vs.\ $I$ at temperatures from 20 to 10~K. 
Each curve is labeled by $T$ in K.  Dashed lines are approximate fits to
the slopes at low current; slope $= p$.  Inset: plot of $p$ vs.\ $T$. 
Dashed line indicates that $p$ crosses 3 at $T=15.6$~K.}
\label{fg:I_V} 
\end{figure}

Next we consider \tctd.  One can see in Fig.~\ref{fg:S_rho} that 
$\rho_{ab}$ rapidly drops by about an order of magnitude at $\sim40$~K,
while the magnitude of the thermopower simultaneously drops to nearly
zero.  It is apparent that \tctd\ is quite sensitive to a magnetic
field applied along the
$c$-axis.  Figure~\ref{fg:I_V}(a) indicates that the transition is also
sensitive to the current used to measure the in-plane resistivity.  In
Fig.~\ref{fg:log_rho}(a), one can see that
$\rho_c/\rho_{ab}$ grows by an order of magnitude; this indicates that
the drop in $\rho_{ab}$ involves purely 2D behavior, with no
communication between the planes.

The sensitivity of \tctd\ to magnetic fields and
current suggests a connection with superconductivity.  In fact, we had
previously attributed the transition to filamentary superconductivity
associated with local variations in hole content \cite{home06}.  There is
a serious problem with this explanation, however: the transition
temperature is higher than the highest bulk $T_c$ (33~K) in the \lbco\
phase diagram \cite{mood88}.

Things get even more interesting when we examine the finite-resistivity
state below \tctd.  The solid lines in
Fig.~\ref{fg:log_rho}(b) are fits to the formula
\begin{equation}
	\rho_{ab}(T) = \rho_n \exp\left(-b/\sqrt{t}\right),
\end{equation}
where $t = (T/T_{\rm BKT})-1$.  This is the predicted \cite{halp79} form
of the resistivity in a two-dimensional superconductor at temperatures
above the BKT transition, $T_{\rm BKT}$,
where true superconductivity is destroyed by phase fluctuations due to the
unbinding of thermally-excited vortex-antivortex pairs
\cite{kost73,bere71}.  This formula is valid only for zero magnetic
field; one expects an activated contribution to $\rho_{ab}$ due to
field-induced vortices.  Instead, we obtain a reasonable fit with
Eq.~(1) by allowing the parameters to be field dependent (see
Table~\ref{tb:params}).  Note that the nominal $T_{\rm BKT}$ at 1~T and
above falls into the regime where $\rho_c\sim0$ in zero field [see inset
of Fig.~\ref{fg:log_rho}(a)] and the resistivity might be dominated by
imperfections of the sample.

\begin{table}[t]
\caption{Values of parameters obtained in fitting Eq.~(1) to the
resistivity data in Fig.~\ref{fg:log_rho}(b).  Numbers in parentheses are
uncertainties in the last digit.}
\begin{ruledtabular}
\begin{tabular}[c]{crlc}%
$\mu_0 H$ & \multicolumn{1}{c}{$T_{\rm BKT}$} &
  \multicolumn{1}{c}{$\rho_n$} & $b$ \\ 
(T) & \multicolumn{1}{c}{(K)} & \multicolumn{1}{c}{(m$\Omega$ cm)} &  \\
\cline{1-4}
0 & 16.3(3) & 0.13 & 2.7(1) \\
1 & 10.0(5) & 0.041(6) & 2.1(1) \\
3 & 3.8(3)  & 0.022(3) & 2.0(1) \\
\end{tabular}%
\end{ruledtabular}%
\label{tb:params}%
\end{table}

In a 2D superconductor, one expects to have a critical current of
zero and
\begin{equation}
V \sim I^p,
\end{equation}
with $p=3$ just below $T_{\rm BKT}$ and growing with decreasing
temperature \cite{minn87}. Figure~\ref{fg:I_V}(b) shows plots of $V$
vs.\ $I$ at temperatures spanning $T_{\rm BKT}=16.3$~K.  We see that on
approaching the transition, $p$ deviates from 1; it grows rapidly below
16~K.  By interpolating, $p$ reaches 3 at $T=15.6$~K, close to the \tbkt\
determined by fitting $\rho_{ab}(T)$.

\begin{figure}[t]
\centerline{\includegraphics[width=2.8in]{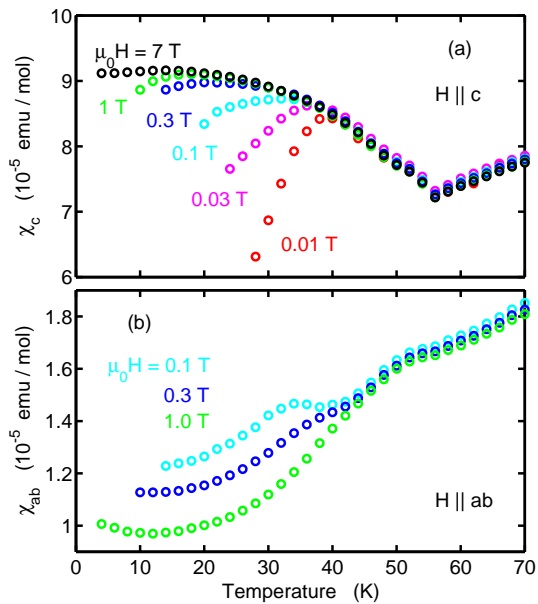}}
\caption{(color online)  (a) $\chi_c$ vs.\ temperature for several
different magnetic fields, as labeled.  (b) $\chi_{ab}$ vs.\ temperature
for several magnetic fields as labeled. Note that, for each applied field,
measurements at temperatures lower than the plotted data range are
unreliable due to excessive noise in this regime, possibly associated
with the unusual state of the sample.}
\label{fg:susc} 
\end{figure}

If we truly have 2D superconducting fluctuations present within the
CuO$_2$ planes below \tctd, then we would expect to see a
weak diamagnetic response in the magnetic susceptibility $\chi_c$, as a
field applied along the $c$-axis should generate an orbital response in
the planes.  In contrast, there should be no diamagnetic response in
$\chi_{ab}$, with the field parallel to the planes.  Figure~\ref{fg:susc}
shows measurements of $\chi_c$ and $\chi_{ab}$ vs.\ temperature on
cooling in field; $\chi$ is reversible for the data ranges shown.  One can
see that the susceptibility is dependent on the magnetic field used for
the measurement.  For ${\bf H}\parallel{\bf c}$, the high-field
susceptibility is dominated by the response of the ordered Cu moments
\cite{huck05b}. Relative to that, we see that there is a weak diamagnetism
below $\sim40$~K that decreases with increasing field, consistent with
superconducting correlations.  For ${\bf H}\perp{\bf c}$, there is no 
diamagnetism, as expected. Instead, $\chi_{ab}$ actually decreases with
increasing field.  This is due to a paramagnetic contribution that
saturates at a field of $\sim1$~T and remains to be understood.   The
weak, anisotropic diamagnetism looks similar to the results of Li {\it et
al.} \cite{li07} in under-doped \lsco. 

The decoupling between the planes in our sample is not perfect, and 
defects are likely to become increasingly relevant as the temperature
decreases.  At twin boundaries, the crystal structure is modified, and a
local coupling might be possible. The statistical distribution of dopant
ions could also lead to local variations.  In zero field, $\rho_c$
starts to decrease below $\sim35$~K [inset of Fig.~\ref{fg:log_rho}(b)],
although the ratio $\rho_c/\rho_{ab}$ remains $>10^4$.  Magnetic
susceptibility measured (after zero-field cooling) in a field of 2 Oe
applied parallel to the planes shows the onset of weak diamagnetism at
28~K, reaching $\sim 1$\%\ of the full shielding response at 10~K.

We see that we have a number of experimental signatures compatible with
2D superconducting fluctuations below \tctd.  The necessary
decoupling of the planes is consistent with the highly anisotropic state
below $T_{\rm co}$.  It appears that $T_{\rm so}$ provides an upper limit
to the onset of 2D superconducting fluctuations.  Furthermore, there are
indications of true 2D superconductivity for $T<T_{\rm BKT}\approx16$~K. 
Theoretically, such behavior requires that the net interlayer Josephson
coupling equal zero; Berg {\it et al.} \cite{berg07} have proposed a
plausible model for frustration of the coupling.


To summarize, we have found that the main impact of stripe
ordering is to electronically decouple the CuO$_2$ planes.  Fluctuating 2D
superconductivity appears below \tctd, with a finite resistivity due to
phase fluctuations.  $\rho_{ab}$ goes to zero at a BKT transition. The
evidence of 2D superconducting correlations indicates that static stripes
are fully compatible with pairing, and we note that the high value of
\tctd\ correlates with the maximum of the antinodal gap at $x=\frac18$
\cite{vall06}. The downside is that stripe order, at least as realized in
\lbco, competes with superconducting phase order.  

We are grateful to S. A. Kivelson, A. V. Chubukov, M. Strongin, and T.
Valla for important suggestions and critical comments.  This work was
supported by the Office of Science, U.S. Department of Energy under
Contract No.\ DE-AC02-98CH10886.


\end{document}